\documentclass[a4paper,twocolumn,superscriptaddress,11pt,accepted=2018-11-16]{quantumarticle}
\pdfoutput=1
\usepackage[utf8]{inputenc}
\usepackage[T1]{fontenc}
\usepackage[english]{babel}
\usepackage{graphicx}  
\usepackage{dcolumn}   
\usepackage{bm}        
\usepackage{amssymb}   
\usepackage{amsmath}
\usepackage{mathtools}
\usepackage[utf8]{inputenc}
 \usepackage[numbers, compress]{natbib}
\usepackage{changes}

\definecolor{myurlcolor}{rgb}{0,0,0.7}
\definecolor{myrefcolor}{rgb}{0.8,0,0}
\usepackage{hyperref}
\hypersetup{colorlinks, linkcolor=myrefcolor,
citecolor=myurlcolor, urlcolor=myurlcolor}

\usepackage{cleveref}
\crefrangeformat{equation}{#3(#1 #4--#5 #2)#6}

\usepackage{comment}
\usepackage{amsthm}

\newcommand{\ignore}[1]{}

 \definecolor{jordi}{rgb}{0.1,0.1,0.5}

\begin{document}
\selectlanguage{english}

\title{Bell correlations at finite temperature}
\date{\today}
\author{Matteo Fadel}\email{matteo.fadel@unibas.ch} \affiliation{Department of Physics, University of Basel, Klingelbergstrasse 82, 4056 Basel, Switzerland} 
\orcid{0000-0003-3653-0030}

\author{Jordi Tura} \email{jordi.tura@mpq.mpg.de} 
\affiliation{Max-Planck-Institut f\"ur Quantenoptik, Hans-Kopfermann-Stra{\ss}e 1, 85748 Garching, Germany}
\orcid{0000-0002-6123-1422}

\maketitle

\begin{abstract}
We show that spin systems with infinite-range interactions can violate at thermal equilibrium a multipartite Bell inequality, up to a finite critical temperature $T_c$.
Our framework can be applied to a wide class of spin systems and Bell inequalities, to study whether nonlocality occurs naturally in quantum many-body systems close to the ground state. Moreover, we also show that the low-energy spectrum of the Bell operator associated to such systems can be well approximated by the one of a quantum harmonic oscillator, and that spin-squeezed states are optimal in displaying Bell correlations for such Bell inequalities.
\end{abstract}
\section{Introduction}

One of the key features of quantum physics, which challenges our every-day intuition, is the existence of nonlocal correlations \cite{Bell64}: Local measurements on a quantum system may lead to correlations whose nature departs entirely from the local hidden variable model (LHVM) paradigm, meaning that such correlations cannot be simulated locally by deterministic strategies, even if assisted by shared randomness \cite{Fine82}.

Since Bell showed that quantum mechanics can produce correlations that do not admit a LHVM description \cite{Bell64}, nonlocal correlations have had a fundamental interest \emph{per sé}. Much work has been devoted to understand and characterize them (see \cite{BrunnerRMP2014} for a review). Moreover, Bell correlations constitute a resource that enables novel device-independent (DI) quantum information processing (QIP) tasks (such as DI quantum key distribution \cite{AcinDIQKD, PironioDIQKD}, DI randomness amplification/expansion \cite{ColbeckRenner2012, PironioNature2010} or DI self-testing \cite{MayersYaoST2004, ColadangeloNatComms2017, SATWAP}), thus providing an extra layer of security in QIP protocols, allowing to reduce their assumptions to a minimum level.

Bell correlations are strongly intertwined with entanglement, perhaps the most significant quantum phenomenon \cite{HHHH}. However, it is well known that entanglement and Bell correlations are inequivalent both in the bipartite \cite{WernerStates} and in the multipartite \cite{AugusiakPRLInequivalence, BowlesPRL2016, AugusiakArxivInequivalenceExtension} cases: Bell correlations are strictly stronger as they can only arise from entangled states, while the converse is not true in general (see e.g. \cite{ReviewLocalModels}).

In the many-body regime, entanglement has proven to be key to the understanding of a plethora of physical phenomena \cite{AmicoRMP2008}. Intensive studies of entanglement have elucidated its important role in modern physics' central topics such as quantum phase transitions \cite{VojtaRepPro2003}, superconductivity at high temperature \cite{NelsonPRL1988}, the holographic principle \cite{RyuPRLHolographic2006} or quantum chemistry reactions \cite{LanyonNatChem2010, SzalayIJQC2015}. Furtermore, entanglement lies at the heart of the tensor network ansatz \cite{VMCTN}, and could even be relevant to the understanding of quantum machine learning models \cite{LiuML2017}.

However, the role played by nonlocal correlations in many-body physics is not so clear yet. Their complex characterization (see \cite{PitowskyBook1989}), which is known to be NP-hard even in the bipartite case \cite{Babai1991}, has posed a barrier to the study of multipartite scenarios. For this reason, the implications of nonlocal correlations in systems consisting of a very large number of parties has been mostly lacking, despite remarkable results have been obtained throughout the years \cite{MerminPRL1990, WernerWolfPRA2001, ZukowskiBruknerPRL2001, BancalPRL2011, ChenPRA2011, BellIneqsGraphStates, BellIneqsGraphStates2, CavalcantiCVBellIneqs, HePRL2009, BellIneqsCVMeasurements, SallesMultilinearContractions}. Interestingly, recent theoretical advances developed in the context of few-body Bell inequalities \cite{SciencePaper, AnnPhys, ThetaBodies, WagnerPRL2017, TuraPRX2017, TIPaper, WangPRL2017, Navascues2D, BaccariPRX2017} have allowed for simple ways to detect these correlations in many-body systems. On the one hand, it has been shown that many-body observables such as the energy of the system could signal the presence of Bell correlations in the system \cite{TuraPRX2017}. On the other hand, routinely measured experimental quantities such as total spin components and higher moments thereof \cite{Sorensen2001, EckertNatPhys}, have been shown to be sufficient to formulate Bell correlation witnesses that can detect nonlocality in many-body systems \cite{SciencePaper, AnnPhys, ThetaBodies, WagnerPRL2017}. With these tools, such correlations have been observed experimentally in a Bose-Einstein Condensate of $480$ atoms \cite{ExperimentBasel} and in a thermal ensemble of $5 \times 10^5$ atoms \cite{ExperimentKasevich}.

In this work, we study whether Bell-nonlocal states  (i.e., quantum states with the potential to produce Bell correlations if one were to perform a proper loophole-free Bell test on them) appear naturally in the low-energy spectrum of physically relevant Hamiltonians. As shown in \cite{SciencePaper, AnnPhys, TuraPRX2017}, partial answers to this question are given, by noticing that the ground state of some spin Hamiltonians can violate a Bell inequality. However, conclusions for more experimentally relevant situations, such as non-zero temperature or systems with infinite-range interactions are still missing.

Here, we show that a many-body spin system at finite temperature can violate a Bell inequality and we quantify this violation. We also show that the low-energy part of the spectrum of the Bell operator corresponding to the optimal measurements of the Bell inequality can be well approximated by a quantum harmonic oscillator, enabling a connection to squeezed states. We observe a critical temperature above which nonlocality cannot be detected with our inequality, raising the question of whether this phenomenon can be observed in other systems and with other inequalities.
This could allow to observe Bell-nonlocal states appearing naturally in solid-state systems cooled below some transition temperature.

\section{Spin model}
\label{sec:spin}
We consider a system of $N$ spins, described by an Ising model with infinite-range interactions. With ${\mathbf{s}}^{(i)} = (s_x^{(i)},\ s_y^{(i)},\ s_z^{(i)})$ denoting the vector of spin operators for the $i$-th particle, the Hamiltonian of the system is
\begin{equation}\label{spin_model}
H_0 = \sum_{i=1}^N {\mathbf{B}}\cdot {\mathbf{s}}^{(i)} + J \sum_{i<j}^N s_z^{(i)} s_z^{(j)} \;,
\end{equation}
where ${\mathbf{B}} = \left(B_x,B_y,B_z\right)$ is the external magnetic field and $J$ the strength of the spin-spin coupling. Since Eq.~\eqref{spin_model} has a preferred $z$ direction, we can assume, without loss of generality, that $B_y = 0$. The case of interest for our study corresponds to $B_z, J > 0$ and $B_x < 0$, which we shall assume throughout the rest of the paper. Since $J>0$, the spin-spin interaction is antiferromagnetic.

Using the definition of collective spin operator ${\mathbf{S}} = (S_x,\ S_y,\ S_z)$, given by 
\begin{equation}\label{coll_spin}
{\mathbf{S}} = \sum_{i=1}^N  {\mathbf{s}}^{(i)} \;,
\end{equation}
we rewrite Eq.~\eqref{spin_model} as
\begin{equation}\label{spin_model_coll}
H_0 = B_x S_x + B_z S_z + \dfrac{J}{2} \left( S_z^2 - \dfrac{N}{4} \right) \;.
\end{equation}
Here, and in the following, we shall not drop the constant term in the Hamiltonian, as it will be important later on.

Hamiltonians of this form, where interactions are of infinite range, are routinely realized in a number of experiments on many-body systems. Examples include Bose-Einstein condensates \cite{Esteve08,Gross10,Riedel10}, atomic ensembles in optical cavities \cite{Leroux10}, and ion crystals \cite{Britton12,Bohnet16}.

We are interested in the limit $N\rightarrow \infty$, for which the energy of $H_0$ (cf. Eq. \eqref{spin_model_coll}) is minimized when the collective spin has the maximum length $S = N/2$ and pointing direction along $x$ (since $B_x<0$). To approximate the low-energy spectrum of $H_0$, we perform the Holstein-Primakoff transformation, which maps the collective spin to a bosonic mode \cite{HolsteinPrimakoff}
\begin{subequations}
\begin{align}\label{HP}
S_x &= (S - a^\dagger a)\;, \\
S_z - i S_y &= \left(\sqrt{2S - a^\dagger a}\right) a \;, \label{eq:HPMinus}\\
S_z + i S_y &= a^\dagger \left(\sqrt{2S - a^\dagger a}\right) \label{eq:HPPlus}\;,
\end{align}
\end{subequations}
with $a^\dagger$, $a$ being respectively the bosonic creation and annihilation operators acting on the Fock space, and satisfying the canonical commutation relation $[ a , a^\dagger ]= 1$. Because we are interested in the low-energy excitations above the ground state, \textit{i.e.} the case where $S\approx N/2$ and $a^\dagger a \approx 0$, we further approximate $\sqrt{2S - a^\dagger a} \approx \sqrt{2S}$ in Eqs. \eqref{eq:HPMinus} and \eqref{eq:HPPlus}. Note that $a^\dagger a \approx 0$ implies we are considering a small number of bosonic excitations. These bosonic modes arise from the Holstein-Primakoff transformation and can be seen as \textquotedblleft fictitious\textquotedblright\ bosonic modes. As it can be seen from \eqref{HP}, they are excitations of the spin system that reduce $S_x$. This allows us to approximate Eq.~\eqref{spin_model_coll} with the expression
\begin{align}\label{spin_model_HP}
H_1 =&  B_x \left(S - a^\dagger a \right) + \dfrac{B_z \sqrt{S}}{\sqrt{2}} \left(a + a^\dagger \right) + \nonumber\\
& + \dfrac{J}{4}\left( S \left(a + a^\dagger \right)^2 - \dfrac{N}{2} \right) \;.
\end{align}
This Hamiltonian is diagonalized by performing the Bogoliubov (squeezing) transformation \cite{Bogoljubov1958, Valatin1958}
\begin{subequations}\label{bogo}
\begin{align}
a &= \cosh(\xi) b + \sinh(\xi) b^\dagger + w\\
a^\dagger &= \cosh(\xi) b^\dagger + \sinh(\xi) b + w^\ast
\end{align}
\end{subequations}
with $\xi$ and $w$ defined as
\begin{equation}
\xi = \dfrac{1}{4} \log\left(\dfrac{B_x}{B_x - J S} \right) \;,\; w = \dfrac{B_z \sqrt{S}}{(B_x - J S)\sqrt{2}} \;,
\end{equation}
and $b^\dagger$, $b$ being newly defined bosonic creation and annihilation operators satisfying the canonical commutation relation $[b, b^\dagger]=1$.
With Eqs.~\eqref{bogo} we express Eq.~\eqref{spin_model_HP} in the diagonal form
\begin{align}\label{HO}
H_1 =& \sqrt{B_x (B_x-J S)} \left( b^\dagger b + \dfrac{1}{2} \right) + \nonumber\\
& + B_x \left( S + \dfrac{1}{2} \right) - \dfrac{J N}{8} + \dfrac{B_z^2 S}{2(B_x-J S)} \;.
\end{align}
In the limit $N\rightarrow\infty$ the last term of Eq.~\eqref{HO} tends to $-B_z^2/(2J)$, leaving us with the Hamiltonian $H$, defined as
\begin{align}\label{Hfinal}
H =& \sqrt{B_x (B_x-J S)} \left( b^\dagger b + \dfrac{1}{2} \right) + \nonumber\\
& + B_x \left( S + \dfrac{1}{2} \right) - \dfrac{4 B_z^2 + J^2 N}{8 J} \; .
\end{align}

Note that the Hamiltonian in Eq.~\eqref{Hfinal} describes a quantum harmonic oscillator of frequency
\begin{equation}
\omega(S) = \sqrt{B_x (B_x-J S)} \;,
\end{equation}
and its energy spectrum is given by
\begin{align}\label{Espectrum}
E(S,n) =& \omega(S) \left( n+\dfrac{1}{2} \right) + \nonumber\\
& +B_x \left( S + \dfrac{1}{2} \right) - \dfrac{4 B_z^2 + J^2 N}{8 J},
\end{align}
where $n \in \{0,1,2,...\}$ and $S\in \{N/2, N/2-1, ...\}$. 
Eq.~\eqref{Espectrum} shows us that the ground state of $H_0$ is approximated by the ground state of $H$, which has $n=0$ and $S=N/2$ (remember $B_x<0$).

It is important to realize that the spectrum in Eq.~\eqref{Espectrum} has two characteristic spacings: In the limit $N=2S\rightarrow\infty$, one is $E(S,n+1) - E(S,n) = \omega(S)\approx\sqrt{- B_x J S}$, and the other is $E(S+1,n) - E(S,n) \approx B_x$. This shows that for large $S$, the low-energy part of the spectrum is dominated by excitations of $n\approx 0$ and different $S$. Note that, for large values of $S$, $E(S, n+1)- E(S,n)$ scales as $\sqrt{S}$, whereas $E(S+1,n)-E(S,n)$ remains approximately constant. This implies that for sufficiently large values of $S$, the latter constant becomes effectively arbitrarily small, and therefore many levels of $S$ become accessible.

\begin{figure}[t]
 \centering
\includegraphics[width=1\columnwidth]{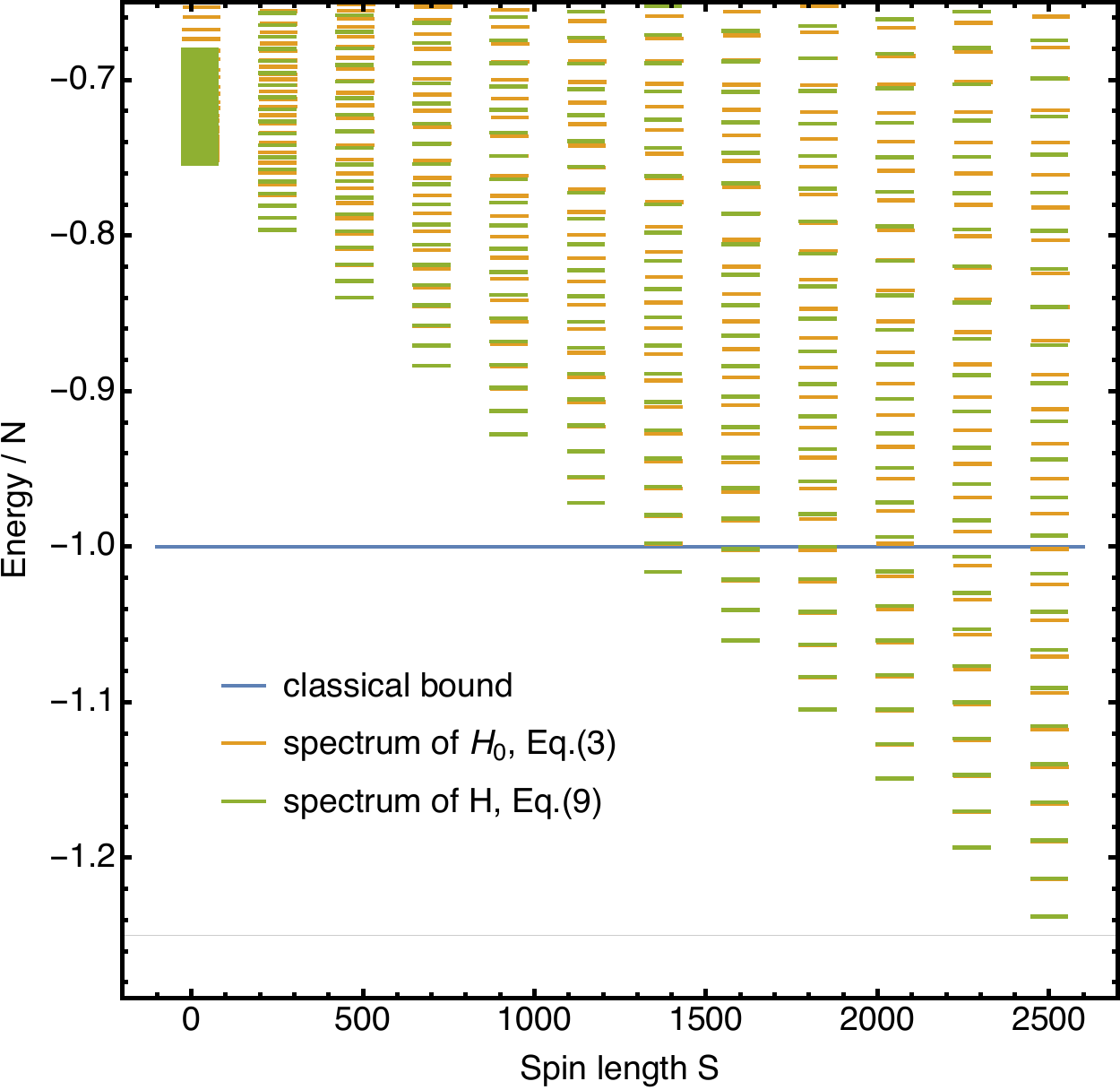}
\caption{\label{fig:spectrum} With $N= 5 \times 10^3$, sample for some values of $S$ of the spectrum of the Hamiltonian Eq.~\eqref{spin_model_coll} compared to the spectrum of the approximated Hamiltonian Eq.~\eqref{Hfinal}, for ${\mathbf{B}} = (-1, 0, \sqrt{3} )$ and $J=6$. Energy is normalized to $N$. Note that for a fixed $S$ the energy levels for the Bell operator are not equally spaced, while they are for the approximated Hamiltonian (harmonic oscillator). The thin gray line at $-5/4$ shows the ground state energy per particle in the limit $N\rightarrow\infty$. The blue line at $-1$ is the classical bound associated to the Bell operator Eq.~\eqref{BellOp}, see Sec.~\ref{sec:BI} for details.}
\end{figure}

For a graphical comparison, we plot in Fig.~\ref{fig:spectrum} the spectrum of the Hamiltonian Eq.~\eqref{spin_model}, and of its approximation Eq.~\eqref{Hfinal}, for $N=5\times 10{^3}$, ${\mathbf{B}} = (-1, 0, \sqrt{3} )$, $J=6$, and some values of $S$ (the reason for this specific choice of parameters will be clear in Sec. \ref{sec:BI}). We can see that low-energy excitations of the two models are very similar. As expected, energy levels for $S\approx N/2$ and $n\approx 0$ are very well described by the Holstein-Primakoff approximation. Note also that this match improves as $N\rightarrow \infty$.

It should be noted that the state yielding the ground energy of the family of Hamiltonians considered is spin-squeezed, as these Hamiltonians are diagonalized via the spin-squeezing transformation (cf. Eq.~\eqref{bogo}). In Sec. \ref{sec:BI} we shall see that the Hamiltonians we consider can be mapped into Bell inequalities. For this reason we conclude that spin-squeezed states are optimal for violating such inequalities, in the sense that they asymptotically (as $N$ goes to infinity) give the maximal quantum violation \cite{AnnPhys}.

To describe the system at finite-temperature, it is crucial to keep track of the degeneracy of the energy levels, originating from the multiplicities $g(S)$ of the blocks with given $S$ (see e.g. \cite{MoroderNJP2012, PhDTura}). This is 1 if $S=N/2$ and otherwise it is given by
\begin{equation}
g(S) = {N \choose N/2 - S} -  {N \choose N/2 - S - 1} \;.
\end{equation}
The derivation of $g(S)$ follows from representation theory results such as the Shur-Weyl duality. We refer the interested reader to \cite{ChristandlThesis, HarrowThesis, AudenaertNotes} for their derivations in quantum information and mathematical contexts.

In the following we will consider two different regimes. The first one is the low temperature regime, where $n\approx0$ and $S\approx N/2$. In this case we can perform the following approximation: For $N\rightarrow\infty$ and $p = N/2 - S \ll N/2$, the leading term in $g(S)$ is given by the factor $N^p/p!$. The second regime of interest corresponds to $S\approx N/4$, and therefore the previous approximation cannot be done. We will discuss how to treat this case in detail in Sections \ref{sec:PF} and \ref{sec:NLvsT}.

\section{Partition function}
\label{sec:PF}
In this section we introduce the standard tools of statistical mechanics that are needed to describe a system at finite-temperature. In particular, expectation values of interest are computed from the partition function of the considered statistical ensemble. As we consider a system in thermal equilibrium at temperature $T$ and with fixed number of atoms $N$, we are in the canonical ensemble.
The canonical partition function for the system described by Eq.~\eqref{Hfinal} is given by
\begin{equation}
Z = \sum_{S,n} g(S) e^{-\beta E(S,n)}
\end{equation}
with $g(S)$ the degeneracy (independent of $n$) of the energy level $E(S,n)$, and $\beta=T^{-1}$ is the inverse temperature. Performing the sum over $n$ leaves us with
\begin{equation}\label{Zexact}
Z =  e^{\beta\left( \frac{4 B_z^2 + J^2 N}{8 J}\right)} \sum_{S} \frac{g(S) e^{-\beta B_x \left( S+\frac{1}{2}\right)}}{2 \sinh \left(\frac{\beta \omega(S)}{2}\right)}\;.
\end{equation}

The argument of the summation is a complicated expression, and for this reason there is no general analytical solution for $Z$. As we will be interested in two specific temperature regimes, we approximate the sum by looking at which terms are relevant. 

In the low temperature regime, $T\approx 0$, the sum over $S$ will be dominated by values of $S\approx N/2$. This can easily be seen e.g. by plotting the terms of the sum in \eqref{Zexact} as a function of $S$. Then one sees that only the terms that are close to $S \approx N/2$ contribute. This observation allows us to set $\omega(S)\simeq\omega(N/2)$, which becomes independent of $S$. Therefore, the denominator of Eq.~\eqref{Zexact} can be taken outside the sum.
Since in this case the approximation $g(S) \simeq N^p/p!$ holds, we can find an analytical expression for the summation over $S$:
\begin{align}
\sum_{S} g(S) e^{-\beta B_x \left( S+\frac{1}{2}\right)} \simeq & \sum_{p} \dfrac{N^p}{p!} e^{-\beta B_x \left( \frac{N}{2}-p+\frac{1}{2}\right)} \\
& = e^{-\beta \frac{B_x}{2} (N+1)} e^{N e^{\beta B_x}} \;.
\end{align}

To summarize, in the low-temperature regime the canonical partition function can be approximated by
\begin{equation}
Z_{T\approx 0} = \dfrac{ e^{\beta\left( \frac{4 B_z^2 + J^2 N}{8 J}\right)} }{2 \sinh \left(\frac{\beta \omega(S)}{2}\right)}  e^{-\beta \frac{B_x}{2} (N+1)} e^{N e^{\beta B_x}} \;.
\end{equation}

The second temperature regime of our interest is the one for which $T\approx 1$, for a reason that will be clear in the following. At this temperature, and for the parameters ${\mathbf{B}}$ and $J$ introduced later, we observe that the terms in the sum of Eq.~\eqref{Zexact} are dominant for $S\approx N/4$. In this situation, we cannot provide an accurate and simple approximation to $g(S)$, and therefore we adopt a different approach. By observing that in the limit $N\rightarrow\infty$ the terms in the summation of Eq.~\eqref{Zexact} follow a Gaussian distribution with mean $S=N/4$, we approximate them in this way. Finally, the sum is replaced by an integral which can be easily computed for a Gaussian. The resulting (approximated) partition function $Z_{T\approx 1}$ is an analytical expression too complicated to be written here (see Appendix), and therefore we will only present some results computed from it.

The crucial quantity of our interest will be the total energy of the system at finite temperature, which is defined as
\begin{equation}\label{meanE}
\langle H \rangle_T = - \dfrac{\partial}{\partial \beta} \log Z.
\end{equation}

In the regime $T\approx 0$ and $N\rightarrow\infty$ we find
\begin{equation}\label{HlowT}
\langle H \rangle_{T\approx 0}  \simeq  \left( \dfrac{4 B_x - J}{ 8} - B_x e^{B_x \beta} \right) N + \dfrac{\sqrt{-B_x J N}}{2\sqrt{2}} \;.
\end{equation}
For the parameters of our interest, ${\mathbf{B}} = (-1, 0, \sqrt{3} )$ and $J=6$, this becomes
\begin{equation}\label{HlowT_old}
\langle H \rangle_{T\approx 0}  \simeq  \dfrac{1}{4} \bigg( \left( 4  e^{-\beta } -5 \right) N  + 2 \sqrt{3 N}  \bigg) \;.
\end{equation}

For these specific parameters, we find in the regime $T\approx 1$ and $N\rightarrow\infty$ the simple expression 
\begin{align}\label{HhighT}
\langle H \rangle_{T\approx 1} \simeq &  \dfrac{1}{64} \bigg( 9 \sqrt{6} \sqrt{N} (2 \beta +2-\log (3)) -  \nonumber\\
& \phantom{=}  - 4 N (3 \beta +16-3 \log (3)) \bigg) \;.
\end{align}

\section{Bell inequality}
\label{sec:BI}

In this section we introduce the class of inequalities with which we consider to detect Bell correlations at finite temperature. We note that our scenario of interest is not the one in which a loophole-free Bell test is performed, but rather, in which one certifies the presence of a Bell-nonlocal state; i.e., a quantum state capable to produce Bell correlations if a proper loophole-free Bell test were performed upon it. We also recall the mapping between multipartite Bell inequalities and quantum Hamiltonians. In particular, we show how the Bell operator associated to the inequalities of our interest is related to the Hamiltonian in Eq. \eqref{spin_model}.

We consider the simplest multipartite Bell experiment, in which $N$ spatially separated observers labelled from $1$ to $N$ hold their shares of a quantum system and have at their disposal two dichotomic measurements each. The measurements are labelled $0$ or $1$, and they can yield outcomes $\pm 1$. We denote the $k$-th measurement performed on the $i$-th party ${\cal M}_{k}^{(i)}$. We are interested in Bell inequalities that involve only symmetric one- and two-body correlators, as they are the simplest to be accessed experimentally. These correlators are defined as
\begin{align}
 \mathcal{S}_{k} & = \sum_{i=1}^N \mathcal{M}_k^{(i)},\\
 \mathcal{S}_{kl} & = \sum_{i\neq j} \mathcal{M}_k^{(i)}\mathcal{M}_l^{(j)},
\end{align}
and they yield Bell inequalities of the following form \cite{SciencePaper, AnnPhys}:
\begin{equation}
\label{eq:PIBI}
 I = \sum_{k} \alpha_k {\cal S}_k + \sum_{k \leq l} \alpha_{kl}{\cal S}_{kl} \leq \beta_C,
\end{equation}
where $\beta_C \in {\mathbbm R}$ is the so-called classical bound and $\alpha_k$, $\alpha_{kl}$ are real parameters.

To illustrate our framework, we shall focus on the multipartite Bell inequality considered in \cite{SciencePaper},
\begin{equation}\label{ineq6}
I = -2\mathcal{S}_0 + \dfrac{1}{2} \mathcal{S}_{00} -  \mathcal{S}_{01} + \dfrac{1}{2} \mathcal{S}_{11} \geq - 2 N,
\end{equation}
even if our method can be straightforwardly adapted to any other inequality of the form of Eq. \eqref{eq:PIBI}.

Because Ineq. \eqref{ineq6} has two settings and two outcomes, it can be minimized with traceless real qubit measurements \cite{TonerVerstraete}. For this reason, without loss of generality, it is enough to consider $\mathcal{M}_0^{(i)} = \cos(\phi)\sigma_z^{(i)}+\sin(\phi)\sigma_x^{(i)}$ and $\mathcal{M}_1^{(i)} = \cos(\theta)\sigma_z^{(i)}+\sin(\theta)\sigma_x^{(i)}$.
As shown in \cite{AnnPhys}, only the angle between $\theta$ and $\phi$ matters, and therefore it is also possible to choose $\phi = \pi - \theta$, which allows us to transform the Bell inequality \eqref{ineq6} into the Bell operator
\begin{align}
\mathcal{B}_0 =& -2 \sin(\theta) \sum_{i=1}^N \sigma_x^{(i)} + 2 \cos(\theta) \sum_{i=1}^N \sigma_z^{(i)} + \nonumber\\
& + 2 \cos(\theta)^2\sum_{i\neq j} \sigma_z^{(i)} \sigma_z^{(j)} \succeq - 2 N \;,
\end{align}
where $A \succeq B$ stands for $A-B \succeq 0$; i.e., $A-B$ being a positive semidefinite operator.
Moreover, in the limit of $N\rightarrow\infty$ it was found \cite{AnnPhys} that the optimal measurement angle is $\theta=\pi/6$. This gives, with the definition $\mathbf{s}^{(i)}=\boldsymbol{\sigma}^{(i)}/2$, the Bell operator
\begin{align}\label{BellOp}
\mathcal{B} =& - \sum_{i=1}^N s_x^{(i)} + \sqrt{3} \sum_{i=1}^N s_z^{(i)} + \nonumber\\
& + 6 \sum_{i < j} s_z^{(i)} s_z^{(j)} \succeq - N \;.
\end{align}
Since $\langle \mathcal{B}\rangle \geq -N$ for all local states, observing $\langle \mathcal{B}\rangle < -N$ witnesses nonlocality.

At this point, we can motivate our choice of the parameters ${\mathbf{B}} = (-1, 0, \sqrt{3} )$ and $J=6$ for Eq.~\eqref{spin_model}. This particular case is of special interest as it results in a spin model whose Hamiltonian is identical to the Bell operator Eq.~\eqref{BellOp}.

Now the idea is simple: since the specific Bell inequality and spin Hamiltonian we considered can be mapped one into the other, the energy of our spin system can be identified to the expectation value of the Bell operator, \textit{i.e.} $\langle \mathcal{B}\rangle_T = \langle H_0\rangle_T$. For low enough temperatures the energy of the system will be lower than the classical bound, implying the presence of Bell correlations. 

\section{Nonlocality \textit{versus} temperature}
\label{sec:NLvsT}

\begin{figure}[t]
\centering
\includegraphics[width=1\columnwidth]{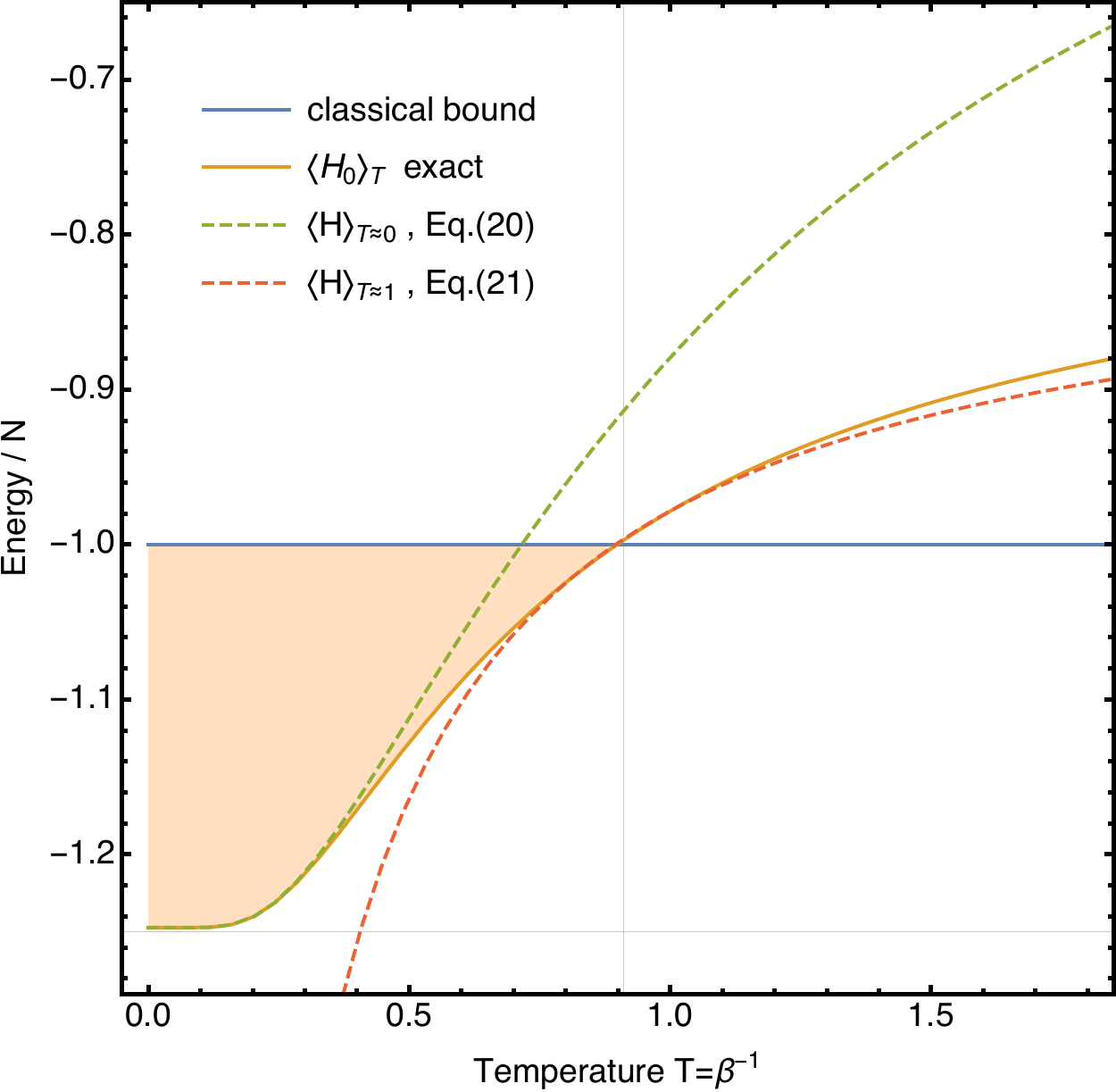}
\caption{\label{fig:energy} With $N=10^5$, mean energy per particle as a function of the temperature. The orange line is the exact mean energy per particle computed numerically (cf. Eq.~\eqref{spin_model} and Eq.~\eqref{meanE}). The green dashed line corresponds to its low temperature approximation (cf. Eq.~\eqref{HlowT}) whereas the red dashed line corresponds to the approximation that is valid around $T\approx 1$ (cf. Eq.~\eqref{HhighT}). The shaded orange region is the one for which nonlocality is detected by considering Eq.\eqref{BellOp}. The thin vertical gray line at $T_c =( \log 3)^{-1}$ corresponds to the value of the critical temperature (cf. Eq.~\eqref{Tc}). The thin horizontal gray line at $-5/4$ corresponds to the maximal quantum violation of Eq.\eqref{BellOp} in the limit $N\rightarrow\infty$ (cf. Eq.~\eqref{maxviol}).
}
\end{figure}

We now investigate the dependence of $\langle \mathcal{B}\rangle_T = \langle H_0\rangle_T$ on the temperature $T$. We plot in Fig.~\ref{fig:energy} the mean energy (cf. Eq.~\eqref{meanE}) per particle as a function of $T$, for $N=10^5$. This modest number of particles allows us to compute numerically the exact partition function Eq.~\eqref{Zexact}, and from it the mean energy per particle (orange line in Fig.~\ref{fig:energy}). Comparing this latter to the classical bound of the Bell operator Eq.~\eqref{BellOp} normalized per particle, $-N/N=-1$ (blue line in Fig.~\ref{fig:energy}), shows a violation of the inequality $\langle \mathcal{B}\rangle_T \geq - N$ up to a temperature of $\approx 0.9$ (see the intersection between the orange and the blue line in Fig.~\ref{fig:energy}).

In the limit $N\rightarrow\infty$, we are interested in finding the critical temperature $T_c$ above which nonlocality is not detected by the inequality we considered.
Naively, from the low temperature approximation Eq.\eqref{HlowT_old} we obtain
\begin{equation}
T_c^{\text{guess}} = \dfrac{B_x}{\log \frac{8+4B_x - J}{8 B_x}}=( \log 4)^{-1} \approx 0.72 \;.
\end{equation}
However, we know from Fig.~\ref{fig:energy} that this estimate is incorrect, as nonlocality can be detected up to $T \approx 0.9  > T_c^{\text{guess}}$. Moreover, already from the fact that $T_c^{\text{guess}} \approx 1$, one sees \textit{a posteriori} that the low temperature approximation is not sufficiently accurate in this regime. For this reason, we adopt the second approximation introduced in Section \ref{sec:PF}, valid in the regime $T \approx 1$.

From the expression of the mean energy given in Eq.~\eqref{HhighT}, we obtain the exact transition temperature which is
\begin{equation}\label{Tc}
T_c =( \log 3)^{-1} \approx 0.91 \;.
\end{equation}
This result is consistent with the validity regime $T \approx 1$ of Eq.~\eqref{HhighT}, and with what we see from Fig.~\ref{fig:energy}.

For experimentally realistic parameters \cite{Riedel10}, $B_x/\hbar = -1 \, \text{s}^{-1}$, $B_z/\hbar = \sqrt{3} \, \text{s}^{-1}$ and $J/\hbar = 6 \, \text{s}^{-1}$, the resulting transition temperature is $T_c \approx 6.95 \; \text{pK}$. Let us remark that such temperature is not the one associated to the motion of the particles, but to the spin degree of freedom.

We emphasize that the critical temperature Eq.~\eqref{Tc} has been derived from considering the specific Bell inequality Eq.~\eqref{ineq6}. Therefore, it is only a lower bound on the temperature that the considered spin system allows us for displaying nonlocal correlations. In fact, there might exist Bell inequalities that are better suited (even though Inequality \eqref{ineq6} has been chosen because it is experimentally accessible), which allow to observe Bell correlations in the regime $T>T_c$. Looking for a better, and maybe an upper, bound on the transition temperature will be addressed in a future work, for example involving multi-setting inequalities \cite{WagnerPRL2017}.

Even if the low temperature approximation Eq.~\eqref{HlowT}  is not suited to describe the regime around the critical temperature, it enables us to study the maximum quantum violation of the inequality we considered. Note that the ground state energy of the system, in the limit $N\rightarrow \infty$, tends to
\begin{equation}\label{maxviol}
\lim_{N\rightarrow \infty} E(N/2,0) = - \dfrac{5 N}{4} \;,
\end{equation}
which coincides with the quantum bound of Ineq.~\eqref{ineq6}, as proven in Ref. \cite{AnnPhys}.

\section{Conclusions}
We have introduced a framework for the study of Bell correlations in a wide class of spin models at finite temperature. This allowed us to show that nonlocal correlations appear naturally in many-body spin systems, when the temperature is below a critical threshold $T_c$. 

Our results show a clear connection between two-body permutationally invariant Bell inequalities and bosonic systems, via the Holstein-Primakoff approximation and the Bogouliubov transformation. In particular, the low-end part of the spectrum of the Bell operator corresponding to the measurements leading to the maximal quantum violation of the Bell inequality, can be well approximated by a quantum harmonic oscillator, thus extremely simplifying the analysis. 

Our work also proves that spin-squeezed states are asymptotically the optimal ones for the class of Bell inequalities \eqref{ineq6}. Therefore, the quantum states that were prepared in previous experiments detecting Bell correlations in many-body systems belonged to that optimal class (spin-squeezed BEC \cite{ExperimentBasel} or thermal ensemble of atoms \cite{ExperimentKasevich}).

Although, for simplicity, we introduced here our framework with a specific Bell inequality (cf. Ineq \eqref{ineq6}), it can easily be generalized.
Of special interest would be to find other inequalities that allow to detect Bell correlations within a larger range of temperatures, and generalizations to inequalities with more measurement settings \cite{WagnerPRL2017} would constitute a good candidate for this task.

We leave as an open problem also the generalization to systems of larger spin. To the best of our knowledge, currently there is no many-body Bell inequality suited to \textit{e.g.} spin-1 systems, and one would need to resort to methods based on SDP hierarchies to plausibly obtain them \cite{ThetaBodies}.

\section{Acknowledgments}
We are grateful to Roman Schmied for the useful discussions and for the help with the numerical code. M. F. was supported by the Swiss National Science Foundation through Grant No 200020\_169591. This project has received funding from the European Union's Horizon 2020 research and innovation programme under the Marie-Sk\l{}odowska-Curie grant agreement No 748549.

\nocite{apsrev41Control}
\bibliographystyle{apsrev4-1}
\bibliography{NL_Temperature}

\section{Appendix}

We give here the expression for the approximated partition function, valid around $T\approx 1$ for $N\rightarrow\infty$. For ${\mathbf{B}} = (-1, 0, \sqrt{3} )$ and $J=6$ we find
\begin{align}
Z_{T\approx 1}& \approx 2^{2 N+3} 3^{\beta  \left(\frac{9}{32} \sqrt{\frac{3}{2}} \sqrt{N}-\frac{3 N}{16}\right)-\frac{3 N}{4}-\frac{5}{6}} \times \nonumber\\
&\times \exp \Bigg[ \left(\frac{3 N}{32}-\frac{9}{32} \sqrt{\frac{3}{2}} \sqrt{N}+\frac{81}{256}\right) \beta^2 +  \nonumber\\
& + \left(N-\frac{9}{16} \sqrt{\frac{3}{2}}\sqrt{N}+\frac{\sqrt{\frac{3}{2}}}{4 \sqrt{N}}-\frac{1}{6}\right) \beta -  \nonumber\\
& - \frac{109}{324 N}+\frac{3}{32} N \log ^2(3)-\frac{5}{18} \Bigg] \;.
\end{align}

\end{document}